\documentclass[12pt,openright,amsmath,amssymb]{revtex4-1}

\usepackage{color}
\usepackage{graphicx}
\usepackage{dcolumn}
\usepackage{bm}
\usepackage{epstopdf}

\newcommand{\be}[1]{\begin{equation}\label{eq:#1}}
\newcommand{\ee}{\end{equation}}
\newcommand{\bea}{\begin{eqnarray}}
\newcommand{\eea}{\end{eqnarray}}

\newcommand{\mb}{\mathbf}
\newcommand{\phd}{\phantom{\dag}}

\newcommand{\up}{^{\phd}}
\newcommand{\noi}{\noindent}
\newcommand{\no}{\nonumber}

\begin{document}
\def\v#1{{\bf #1}}

\title{Alternative paths to realize Majorana Fermions in Superconductor-Ferromagnet Heterostructures}

\author{G. Livanas$^1$}
\author{M. Sigrist$^2$}
\author{G. Varelogiannis$^1$}
\affiliation{$^1$Department of Physics, National Technical University
of Athens, GR-15780 Athens, Greece\\
$^2$Institut f\"{u}r Theoretische Physik, ETH-Z\"{u}rich, CH-8093 Z\"{u}rich, Switzerland}

\begin{abstract}
\textbf{A fundamental obstacle for achieving quantum computation is local decoherence.
One way to circumvent this problem rests on the concepts of topological quantum computation
using non-local information storage, for example on pairs of Majorana fermions (MFs).
The arguably most promising way to generate MFs relies at present on spin-triplet p-wave states of superconductors (SC), which are not abundant in nature, unfortunately.
Thus, proposals for their engineering in devices,
   usually via proximity effect from a conventional SC into materials with strong spin-orbit coupling (SOC), are intensively investigated nowadays.
   Here we take an alternative path, exploiting the different connections between fields based on a quartet coupling rule for fields
   introduced by one of us,
   we
demonstrate that, for instance, coexisting Zeeman field with a charge current would provide the conditions to induce p-wave pairing in the presence of singlet superconductivity. This opens new avenues for the engineering of
robust MFs in various, not necessarily (quasi-)one-dimensional, superconductor-ferromagnet heterostructures, 
including such motivated by recent pioneering experiments that report MFs,
in particular, without the need of any exotic materials with special structures of intrinsic SOC.}

\end{abstract}

\maketitle






Majorana particles are their own anti-particles \cite{Majorana,Wilczek} each comprising half of a fermion such that widely separated pairs of Majorana states constitute nonlocal fermionic states immune to local decoherence
ideal for building hardware elements for topological quantum computation \cite{Kitaev1,Freedman,Nayak,Alicea}.
Spin-triplet p-wave states of superconductors (SC) are known to be  
potential hosts of MFs 
although these are
rarely intrinsic states of materials \cite{SigristUeda,Read,Kitaev}.
In fact, zero-energy Majorana states have been shown on toy models, to emerge at the edges of
spinless one-dimensional p-wave SC wires \cite{Kitaev} and in vortex cores
of certain two-dimensional chiral $p_x+ip_y$ SC states \cite{Read,Ivanov}.

Given the rarity of convenient p-wave SC in nature,
numerous proposals have been put forward for their quantum engineering in devices involving 
conventional SC instead \cite{Fu,LeeProposal,Nagaosa,Sau,Oppen,AliceaPRB2010,Qi,Flesh,Morp,Kotetes}.
Especially, quantum engineering procedures of relevant for MF generation
effective \textit{p}-wave SC fields from conventional SC in combination with
strong SOC materials like Rashba semiconductors \cite{Sau,Oppen,AliceaPRB2010}
or topologic insulators \cite{Fu,Nagaosa}, have been implemented with impressive progress
\cite{Mourik,ExpVortex}.

The most striking and direct experimental evidence of MFs was,
however, reported by scanning tunneling microscopy at the edges of ferromagnetic (FM) Fe wires placed on the [110] surface of SC Pb \cite{Nadj}.
A convincing explanation of this remarkable phenomenon in terms of a FM atomic chain
in proximity with a SC that exhibits strong
intrinsic Rashba SOC has been proposed
\cite{Nadj,Bernevig,Brydon}. If intrinsic Rashba SOC is so strong on the SC Pb surface then
an eventual isolated SC Pb wire with an in-wire field
could exhibit at the edges MFs as well, the same could be true at the cores of vortices on eventual
SC Pb films.

Here we take an alternative path. Exploring the different connections
between the relevant fields based on the \emph{quartet coupling rule} for fields introduced by one of us \cite{GV},
we show that appropriate p-wave SC fields and robust MFs may be induced from singlet SC states
in the presence of FM and supercurrents, 
without the need to assume any intrinsic Rashba SOC.
Our findings not only provide a groundbreaking perspective on these experiments
\cite{Nadj}, they unlock potentially a plethora of related unexplored paths
for the quantum engineering of MFs in SC/FM devices in which intelligent combinations of 
currents and fields play the key role.
As a typical example, we propose 
a versatile trilayer SC/FM/SC device structure 
that can produce MFs through the same quartets mechanism,
illustrating thus how our approach opens new avenues
for the controllable quantum engineering of robust MFs in
SC/FM heterostructures that may involve trivial materials and may not even need to be quasi-one-dimensional
thanks to the directionality of currents. 

Inspired by the experiments we start with the presentation of an alternative device setup
(see Figure 1a) to induce \textit{p}-wave superconductivity and MF using a Zeeman field (FM) and a perimetric supercurrent without relying on intrinsic Rashba SOC.
In order to demonstrate the functioning of our design we introduce here a simple model of a one-dimensional FM nano-wire embedded in the surface of a conventional SC, described by the 2D Hamiltonian ${\cal
H}=\sum_{\bm{i},\bm{j}}\Psi_{\bm{i}}^{\dag}H_{\bm{i},\bm{j}}\Psi_{\bm{j}}$ with the necessary and sufficient ingredients depicted in Figure 1a.

\bea
H_{\bm{i},\bm{j}}&=& t f_{\bm{i},\bm{j}} \tau_3+(\mu_{\bm{i}}\tau_3 -\tau_3 \bm{h}_{\bm{i}} \cdot \tilde{\bm{\sigma}}+\Delta_{\bm{i}}\tau_2\sigma_2)\delta_{\bm{i},\bm{j}} + \bm{J}_{\bm{i}} \cdot \bm{g}_{\bm{i},\bm{j}}\,,
\label{eq:HMF}
\eea

\noi where the Nambu spinor $\Psi_{\bm{i}}^{\dag}= \left(\psi_{\bm{i},\uparrow}^{\dag},\psi_{\bm{i},\downarrow}^{\dag},\psi_{\bm{i},\uparrow}\up ,\psi_{\bm{i},\downarrow}\up \right)$ is referring to the electronic states on lattice site $\bm{i}$. The Pauli matrices $\bm{\tau}$ and $\bm{\sigma}$ act on particle-hole and spin space, respectively.  The electrons move via nearest-neighbour hopping described by the connection matrix $ f_{\bm{i},\bm{j}} = \delta_{\bm{j},\bm{i} \pm \bm{x}}+\delta_{\bm{j},\bm{i} \pm \bm{y}}$ where $ \bm{x} $ and $ \mb{y} $ are in-plane unit vectors,
with a hopping integral $t$. The local chemical potential is denoted by $\mu_{\bm{i}}$, $\bm{h}_{\bm{i}}$ the local vector Zeeman field whereby in our Nambu spinor representation the spin operator is expressed through $ \tau_3 \tilde{\bm{\sigma}}= \tau_3 (\sigma_1,\tau_3\sigma_2,\sigma_3)$. Moreover, we introduce the
pairing field $\Delta_{\bm{i}}$ for the conventional SC phase. A further key element is the current $ \bm{J}_{\bm{i}} = (J_{\bm{i}}^x$,$J_{\bm{i}}^y )$ with the corresponding connection matrices are given by $ \bm{g}_{\bm{i},\bm{j}} = (g_{\bm{i},\bm{j}}^{x}, g_{\bm{i},\bm{j}}^{y}) = (\pm i\delta_{\bm{j},\bm{i} \pm \bm{x}}, \pm i\delta_{\bm{j},\bm{i} \pm \bm{y}} )$.

\begin{figure}[t]
\includegraphics[scale=0.31]{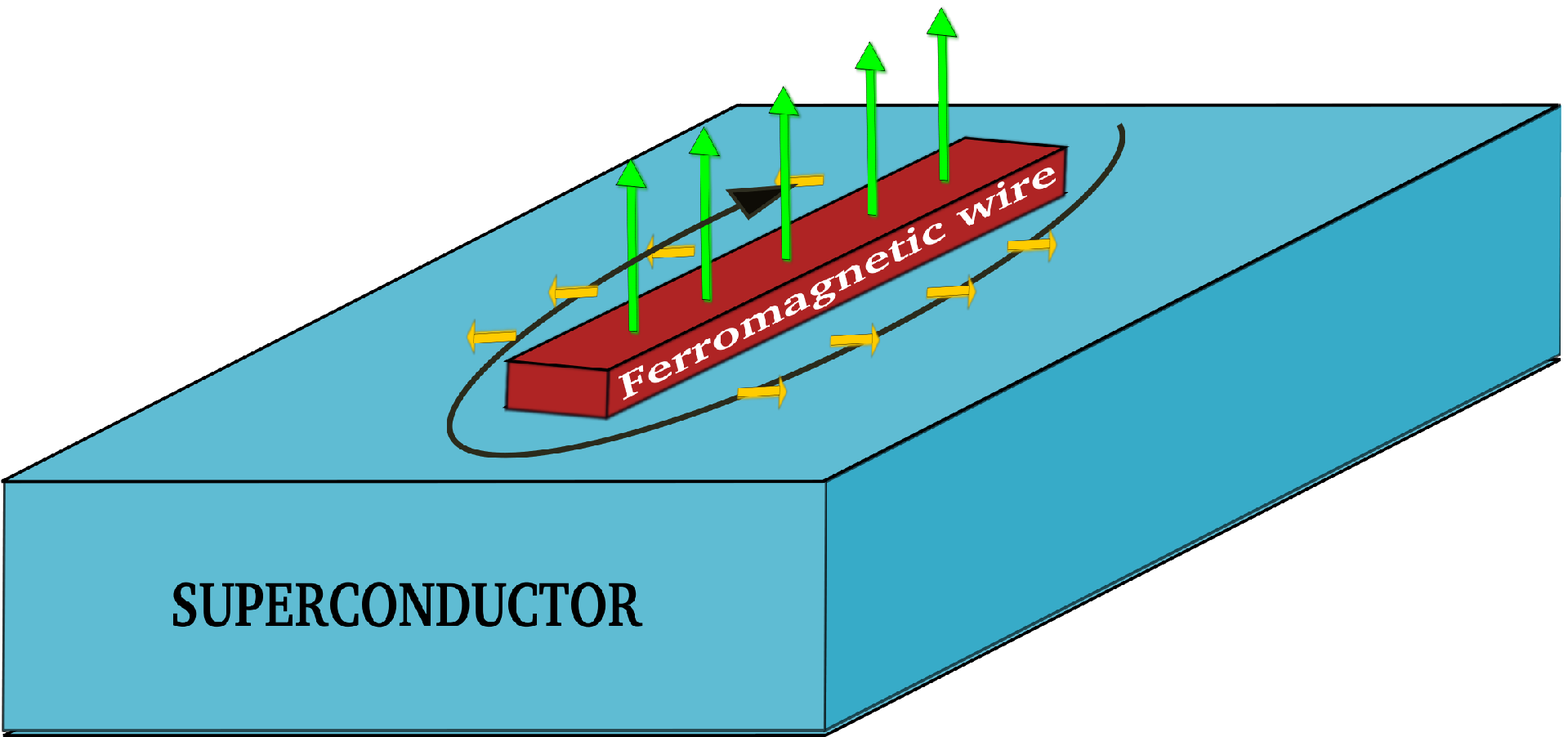}
\includegraphics[scale=0.31]{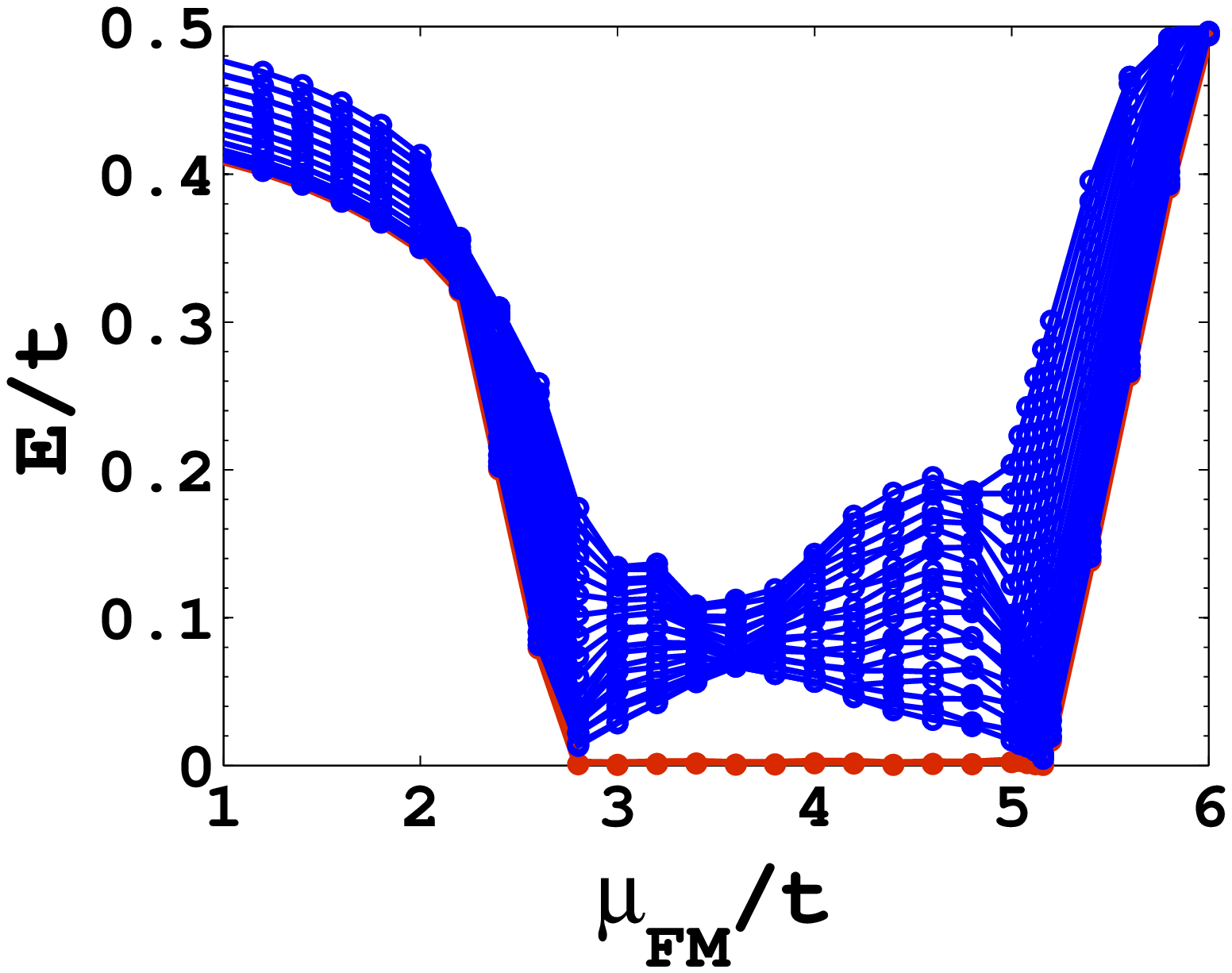}
\includegraphics[scale=0.28]{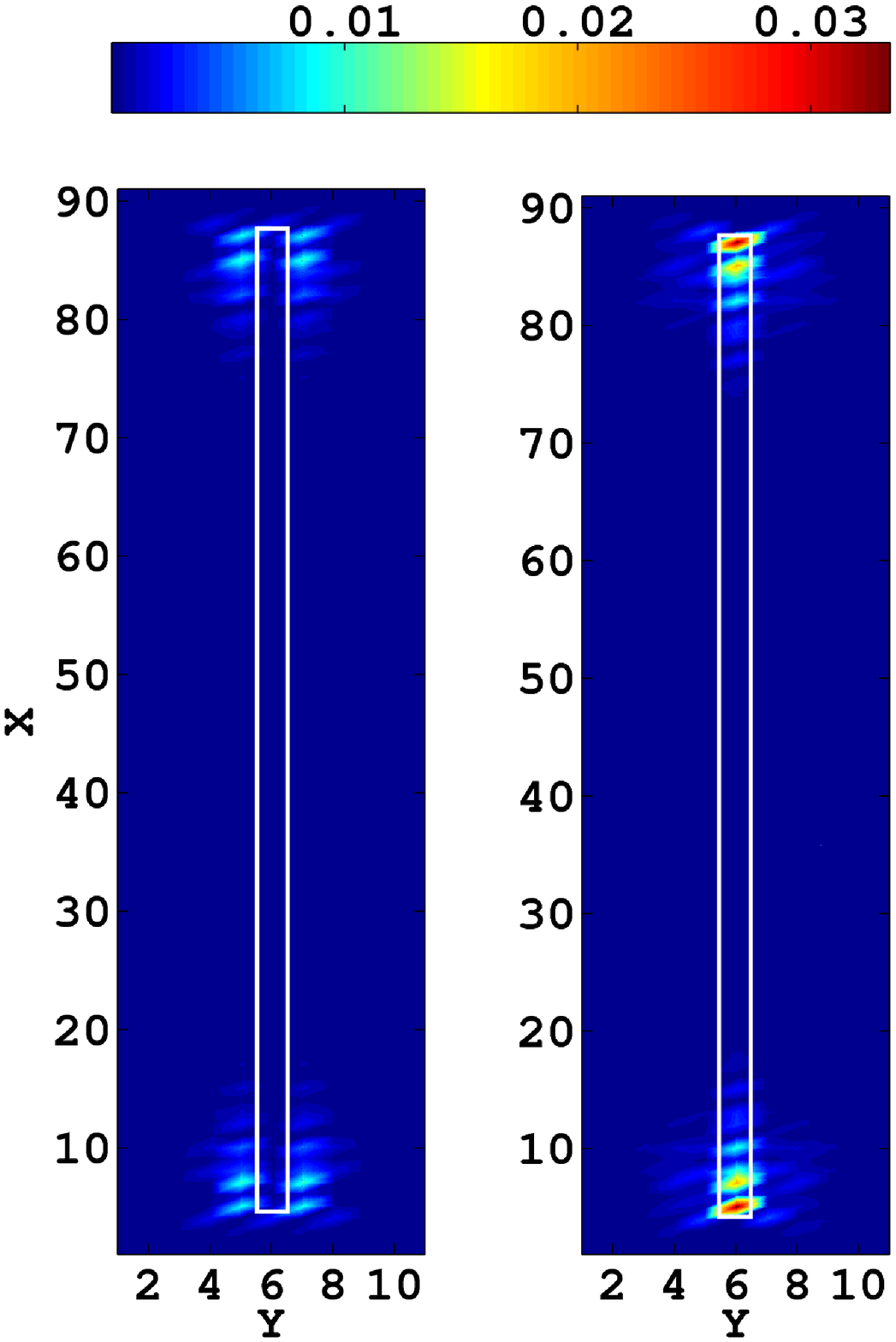}
\includegraphics[scale=0.28]{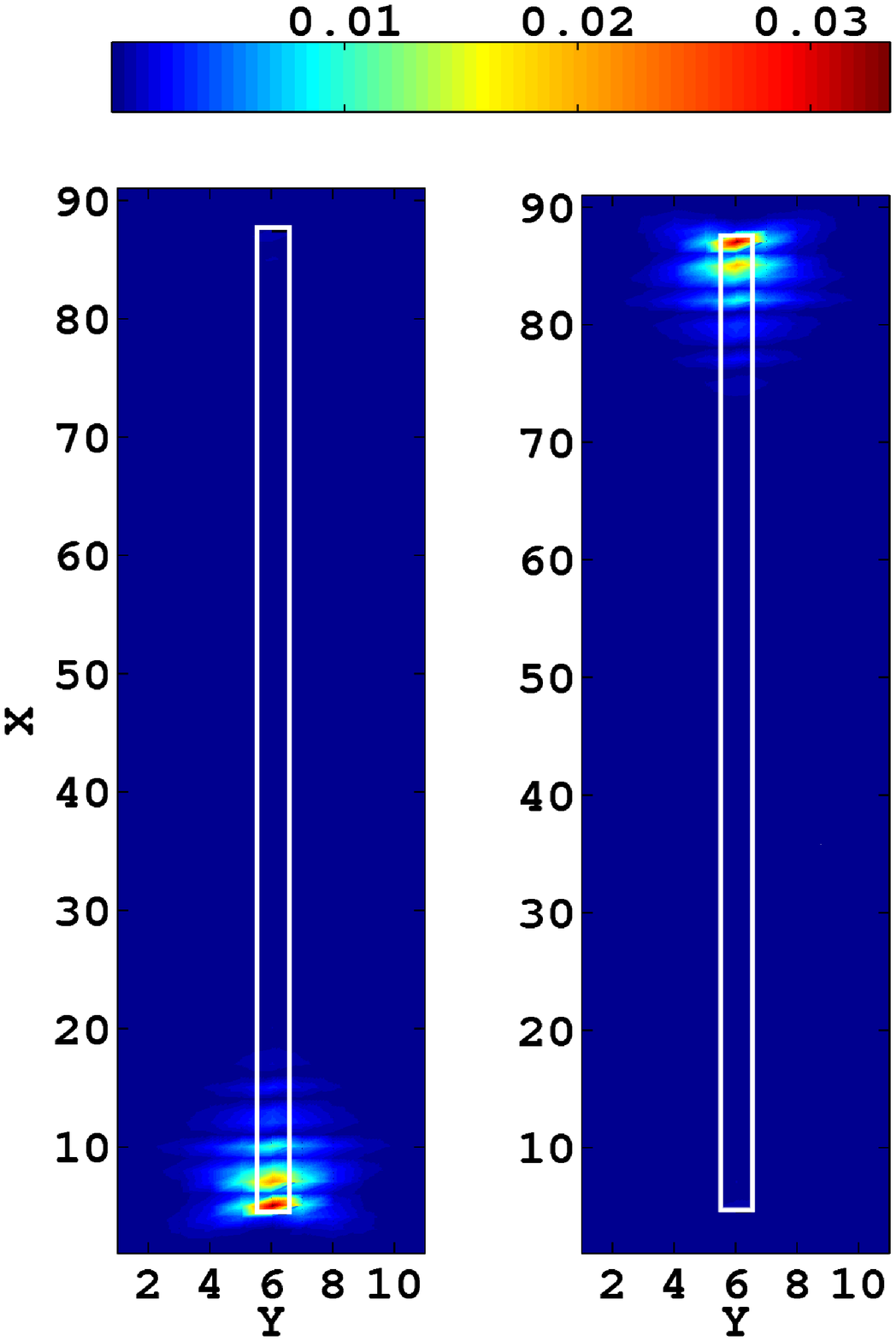}
\caption{\textbf{$\mid$ Heterostructure inspired by the experiment \cite{Nadj}. a,}
One-dimensional FM wire with perpendicular polarization (green arrows) embedded on the surface of a singlet SC,
a supercurrent in the proximity of the wire flowing around it (black arrow) and a small unscreened in plane field component (yellow arrows) in the proximity of the wire. Results remain identical if the sign of the supercurrent and/or the sign of the perpendicular field and/or the sign of all in plane fields is flipped. \textbf{b,} Typical low energy excitation spectrum of Hamiltonian Eq. \ref{eq:HMF} that contains only the ingredients depicted in \textbf{a} with $\Delta=1$ and $\mu_{SC}=0$ for the SC region, $h_z=4$ in the FM wire, $|h_y|=0.4$ and $|J|=0.2$ (all in $t$ units), as a function of the chemical potential in the FM wire $\mu_{FM}$. With red line is highlighted the lowest eigenenergy of the system which pins to zero in the non-trivial topological phase emerging approximately for
$h_z-1.2t<\mu_{FM}<h_z+1.2t$. \textbf{c,} The spin down (left) and spin up (right) parts of the wave function $|\Psi|^2$ corresponding to the lowest eigenenergy of the system in the topologically non-trivial phase for $\mu_{FM}=h_z=4$.
\textbf{d,} The same wave function expressed in the Majorana basis (Supplement II) reveals the two  Majorana fermions localized at the edges of the FM wire. The white rectangle defines the FM wire.} \label{fig:Maj1}
\end{figure}

The setup of our device, as depicted schematically in Figure 1a,
requires that the Zeeman field (magnetic moment) on the FM points along $ z $-axis
(perpendicular to the SC surface) and tilts on adjacent sites perpendicular to the wire ($ y $-direction).
The onsite pairing field $ \Delta_{\bm{i}}$ is constant over the SC region.
The current flows adjacent to FM wire perimetrically and might be considered
as the screening current to the magnetization of FM wire. We use a different
chemical potential for the FM wire ($ \mu_{FM} $) and the SC region ($ \mu_{SC} $).

The straightforward numerical calculations of this model yield a quasiparticle (QP)
spectrum as presented in Figures 1b-d. We observe that a pair of zero energy QP
states appear in the range of $ \mu_{FM-} < \mu_{FM}  < \mu_{FM+} $ with $ \mu_{FM \mp} \approx h_z \mp 1.2 t$, respectively,
for the parameters used (see caption of Figure 1) and indicate the range in which the FM wire would be metallic in the normal state.

The boundaries $\mu_{FM\mp}$ correspond to topological transitions signalled by the closing of the QP gap as seen in Figure 1b. Thus, the topological transitions at $ \mu_{FM \pm} $ coincide essentially with Lifshitz transitions in the electronic bands of the FM wire. Note that the parameters in our numerical treatment imply no overlap of the up and down spin bands.
The QP
wave function of the particle-hole symmetric eigenstates at zero energy is displayed in Figure 1c for $ \mu_{FM} = h_z $. We observe localized states at the two ends of the FM wire, whereby the left (right) panel corresponds to spin down (up) components. These bound states correspond to a pair of MFs as is confirmed by Figure 1d depicting the wave function
in the Majorana basis (see Supplementary Material), one MF on each side.

The origin of this behavior lies in the interplay between the different fields cooperating in the Hamiltonian and can be understood with the scheme of the \emph{quartet rules} put forward by one of the authors \cite{GV}. According to these rules \emph{four} fields (operators) form a \emph{quartet}, if their matrix representations $\widehat{A},\widehat{B},\widehat{C}$ and $\widehat{D}$ obey the relation: $\widehat{A}\widehat{B}\widehat{C}\widehat{D}=\pm\widehat{1}$ \cite{GV}. As a consequence, the presence of any set of three members of a quartet implies that the missing fourth member is intrinsically generated, a phenomenon
named the \emph{quartet rule coupling} between the fields \cite{GV}.
 For example, the combination of charge and spin density wave (CDW and SDW) together with a chemical potential ensuring electron-hole asymmetry can give rise to a ferromagnetic spin polarization,
 important in the context of colossal magnetoresistance \cite{PRLCMR}. Another quartet case has been considered for unconventional superconductors with $ d$-wave pairing combined with a SDW state which in conjunction with electron-hole asymmetry yields a so-called staggered $\pi$-triplet superconducting phase \cite{JoP}, as might be realized in the puzzling high-field low-temperature Q-phase of CeCoIn$_5$\cite{PRLCeCoIn}.

Two such quartets are relevant in our model, specially suitable for engineering of MFs: quartet A composed of charge current, Zeeman field, electron-hole asymmetry and antisymmetric SOC and quartet B with charge current, Zeeman field, conventional singlet SC and \emph{p}-wave triplet SC. Both quartets share the first two fields, but differ in the other two. We use the basic symmetries inversion $ {\cal I} $,  time reversal $ {\cal T} $ and their combination $ {\cal R} = {\cal I} {\cal T} $ to characterize the fields of the quartets as being even ($+$) or odd ($-$) (see table).  In terms of these symmetries electron-hole asymmetry and conventional SC behave equivalently as well as the pair SOC and triplet SC. In case A the quartet rule implies that in a system with electron-hole asymmetry the presence of a charge current $ \bm{J} $ and a Zeeman field $ \bm{h} $ induces SOC of the kind $(\hat{\bm{J}} \cdot \bm{g}_{\bm{i},\bm{j}})( \hat{\bm{h}} \cdot \tilde{\bm{\sigma}})$ with $\hat{\bm{J}},\hat{\bm{h}}$ unitary vectors along  $\bm{J}, \bm{h} $, as is verified within our model and displayed in Figure 2a. In the very same way we see that charge current, Zeeman field and conventional SC drives a spin triplet p-wave component with the real-space structure $  \hat{\bm{J}} \cdot \bm{g}_{\bm{i},\bm{j}} \tau_1(i \sigma_2)( \hat{\bm{h}} \cdot \tilde{\bm{\sigma}}) $ (Figure 2b).

A detailed analysis of the numerical results on Hamiltonian (1) provides insight into the key role of quartet rule coupling
between fields. Besides the creation of the spin triplet component $ \Delta^p_y  \bm{g}_{\bm{i},\bm{j}} \cdot \bm{\hat{x}}  \tau_2 $ through the presence of charge current, Zeeman field $ h_y =\bm{h} \cdot \bm{\hat{y}} $ and the spin-singlet pairing component, the Zeeman field component $ h_z = \bm{h} \cdot \bm{\hat{z}} $ combines with the spin-triplet pairing field $   \Delta^p_y\bm{g}_{\bm{i},\bm{j}} \cdot \bm{\hat{x}} \tau_2 $ and particle-hole asymmetry to induce
$ \Im \Delta^p_x  \bm{g}_{\bm{i},\bm{j}} \cdot  \bm{\hat{x}}  \tau_2 \sigma_3 $ where $\Delta^p_y $($\Im \Delta^p_x $) are even(odd) under time-reversal.
This results from the quartet D discussed in Supplement I.

This combination of triplet pairing fields eventually constitutes
the basis of the Kitaev spinless model \cite{Kitaev}. Based on this
it is also possible now to establish qualitatively the \emph{character}
of the topological phase transition (TPT) suggested by Figure 1b, using an
effective 1D Hamiltonian for the FM wire that contains all induced fields,

\bea
&&{\cal H}_{FM}^{eff}=\sum_{i}\Psi_{i}^{\dag} \left [  \left (t'f_{i,j}^{x} +\mu_{FM}\delta_{i,j} \right )\tau_3 \right . - h_z\delta_{i,j}\tau_3\sigma_3 \no \\
&& +\left.  \Delta' \delta_{i,j}\tau_2\sigma_2  + g_{i,j}^{x}\left (\alpha_y \tau_3\sigma_2 +\Delta^p_y\tau_2 +  \Im \Delta^p_x\tau_2\sigma_3 \right) \right ]\Psi_{j}\up \label{eq::1dRHam} \,.
\eea

\noi with $ t' $ the renormalized hopping matrix element \cite{Peng} with $f_{i,j}^{x} = \delta_{j,i \pm 1} $, $\Delta'$ the singlet pairing component induced by proximity and $ \alpha_y $ the effective SOC appearing through the quartet rule combining charge current, Zeeman field and electron-hole asymmetry \cite{GV}.

Hamiltonian Eq. \ref{eq::1dRHam} belongs to the chiral BDI symmetry class (Supplement III) which for 1D accepts a strong integer $\Bbb{Z}$  topological invariant \cite{Schnyder}. The system is in a non-trivial topological phase with a single pair of zero energy Majorana modes, when $|2t' - \sqrt{(h_z)^2-\Delta'^2}|< |\mu_{FM}| <|2t' + \sqrt{(h_z)^2-\Delta'^2}|$ (Supplement III) that identifies the chemical potential range for which a single energy band is partially occupied. We conclude that the non-trivial topological region in Figure 1b indicates $t'\approx 0.6t$ and is almost symmetric with respect to $\mu_{FM}=h_z=4$ because $\Delta'$ is rather small.

\begin{table}
\begin{tabular}{|l|c|c|c||l|c|c|c|}
\hline
 Quartet A &    ${\cal I}$ &  ${\cal T}$ &  ${\cal R}$ & Quartet B &   ${\cal I}$ &  ${\cal T}$ &  ${\cal R}$   \\
 \hline\hline
 charge current &  $-$ &  $-$ &  $+$  &  charge current & $-$ & $-$ & $+$ \\
\hline
Zeeman field &  $+$ &  $-$ &  $-$  &  Zeeman field &  $+$ &  $-$ &  $-$   \\
\hline
electron-hole asymmetry & $+$ & $+$ & $+$  & conventional SC & $+$ & $+$ & $+$    \\
\hline
spin-orbit coupling &  $-$ &  $+$ & $-$  &  triplet $p$-wave SC & $-$ & $+$ & $-$  \\
\hline
\end{tabular}
\label{Tab1}
\end{table}

 \begin{figure}[t]
\includegraphics[scale=0.25]{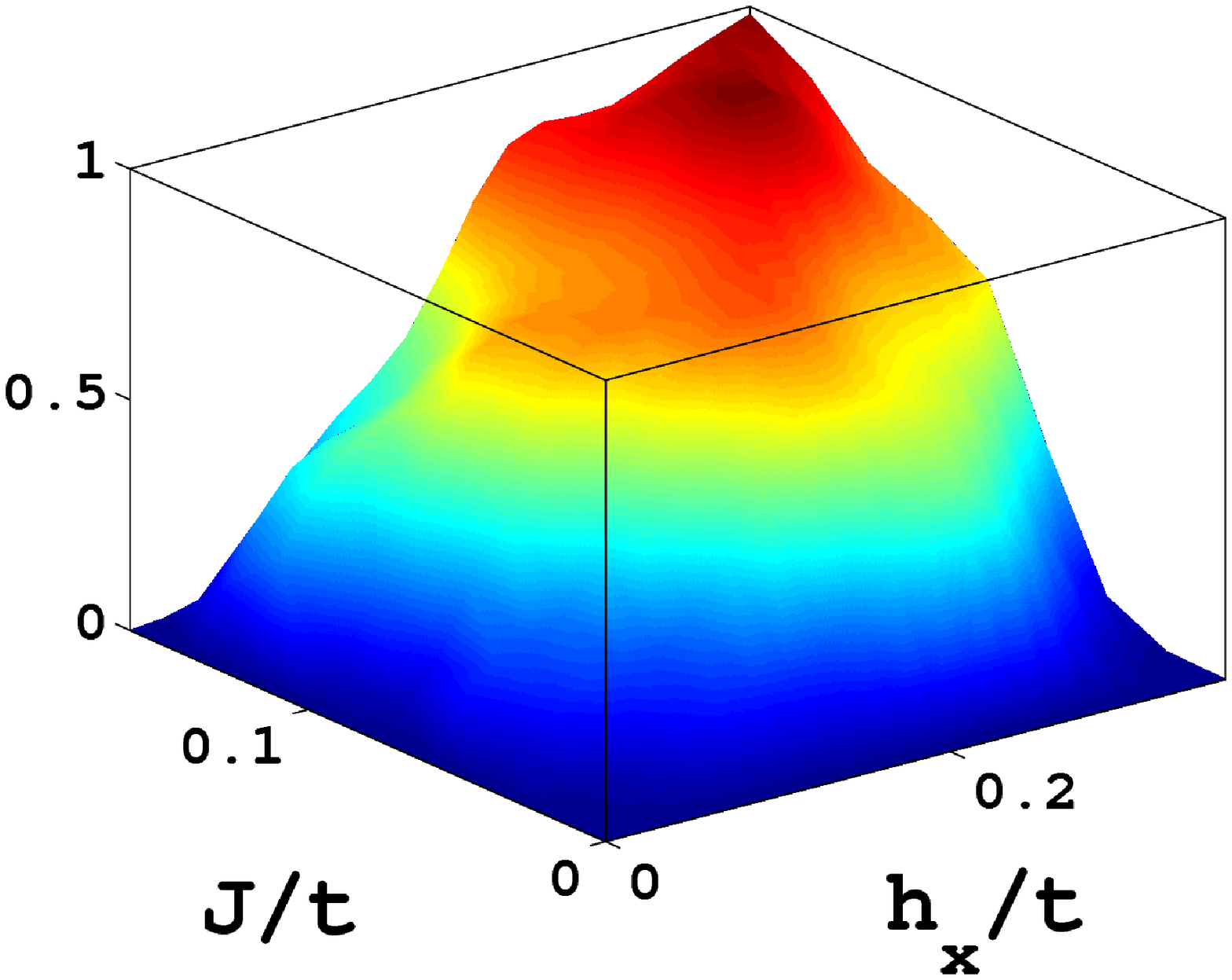}
\includegraphics[scale=0.25]{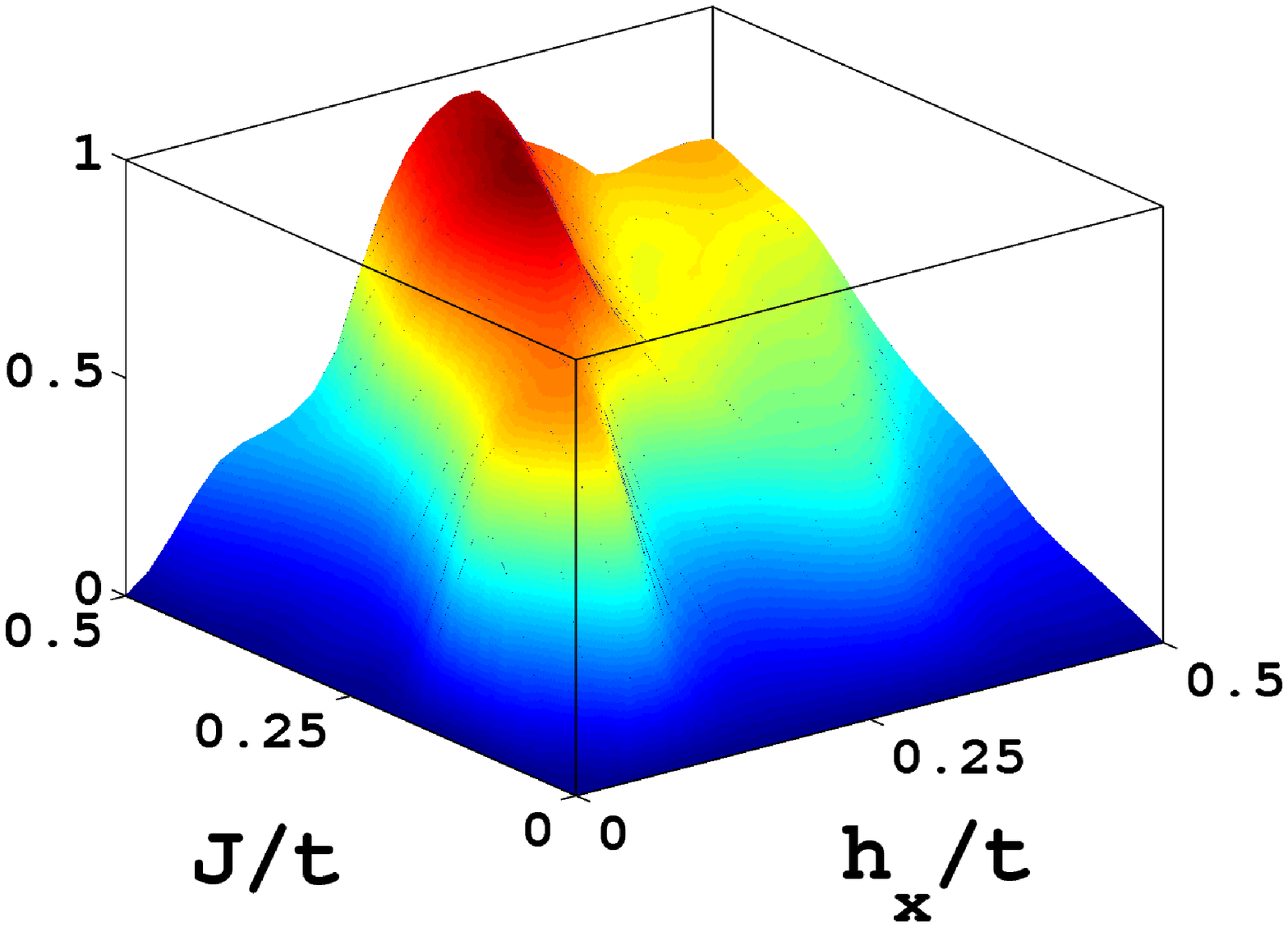}
\caption{\textbf{$\mid$ Quartet rule coupling \cite{GV} for quartets A and B. a,} Induced spin-orbit-coupling (SOC) normalized to its maximal value
as a function of the charge current and the Zeeman field in the presence of finite chemical potential producing particle-hole asymmetry.
\textbf{b,} The same for induced \emph{p}-wave superconductor (SC) in the presence of conventional s-wave superconductor.
Note that only when \emph{both}
the current and the Zeeman field are non zero, the quartet rule coupling applies and we have
the induced SOC and \emph{p}-wave SC fields confirming quartets A and B respectively (see Table and Supplement I).}\label{fig:1quartetsAB}
\end{figure}

\begin{figure}[t]
\includegraphics[scale=0.31]{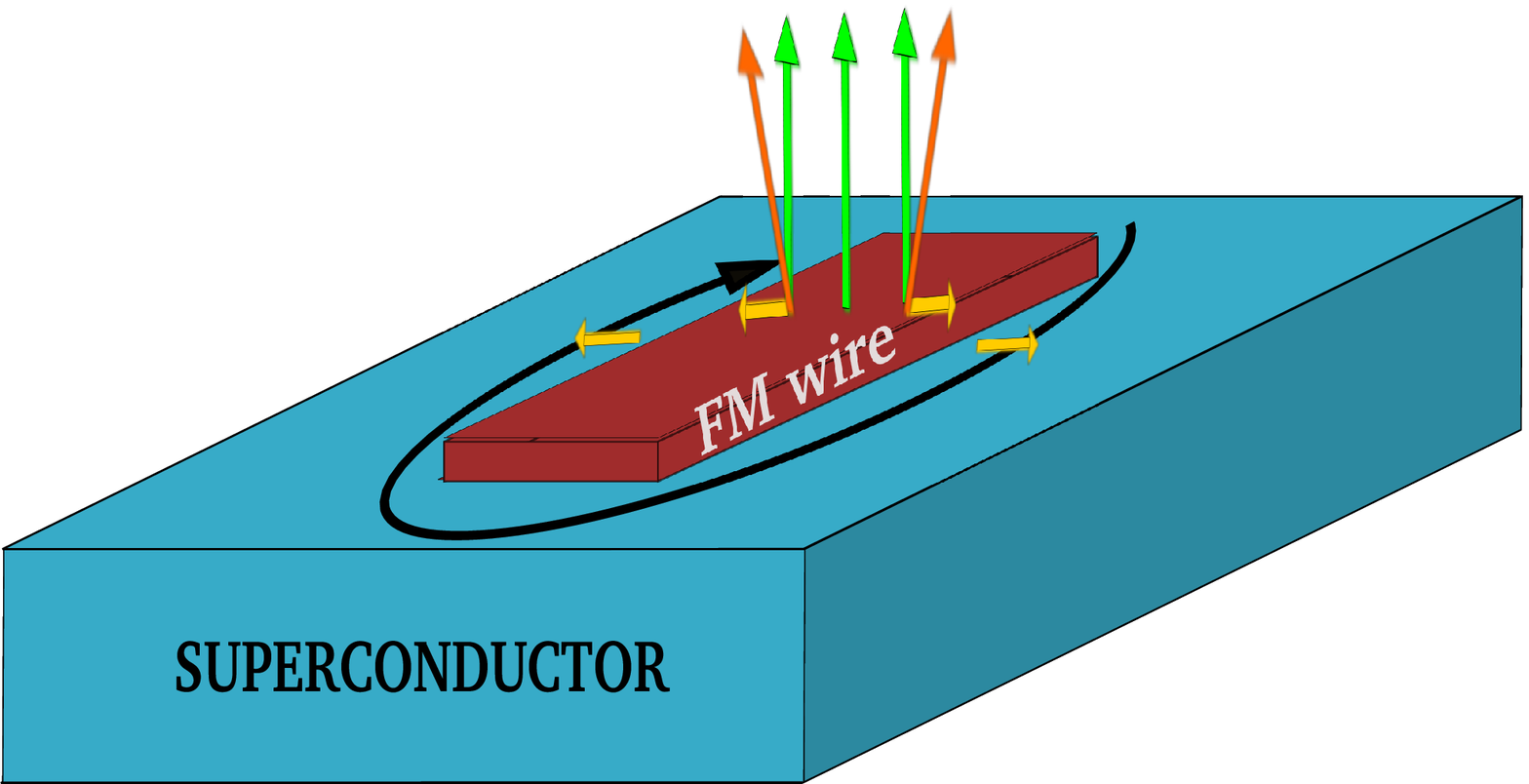}
\includegraphics[scale=0.31]{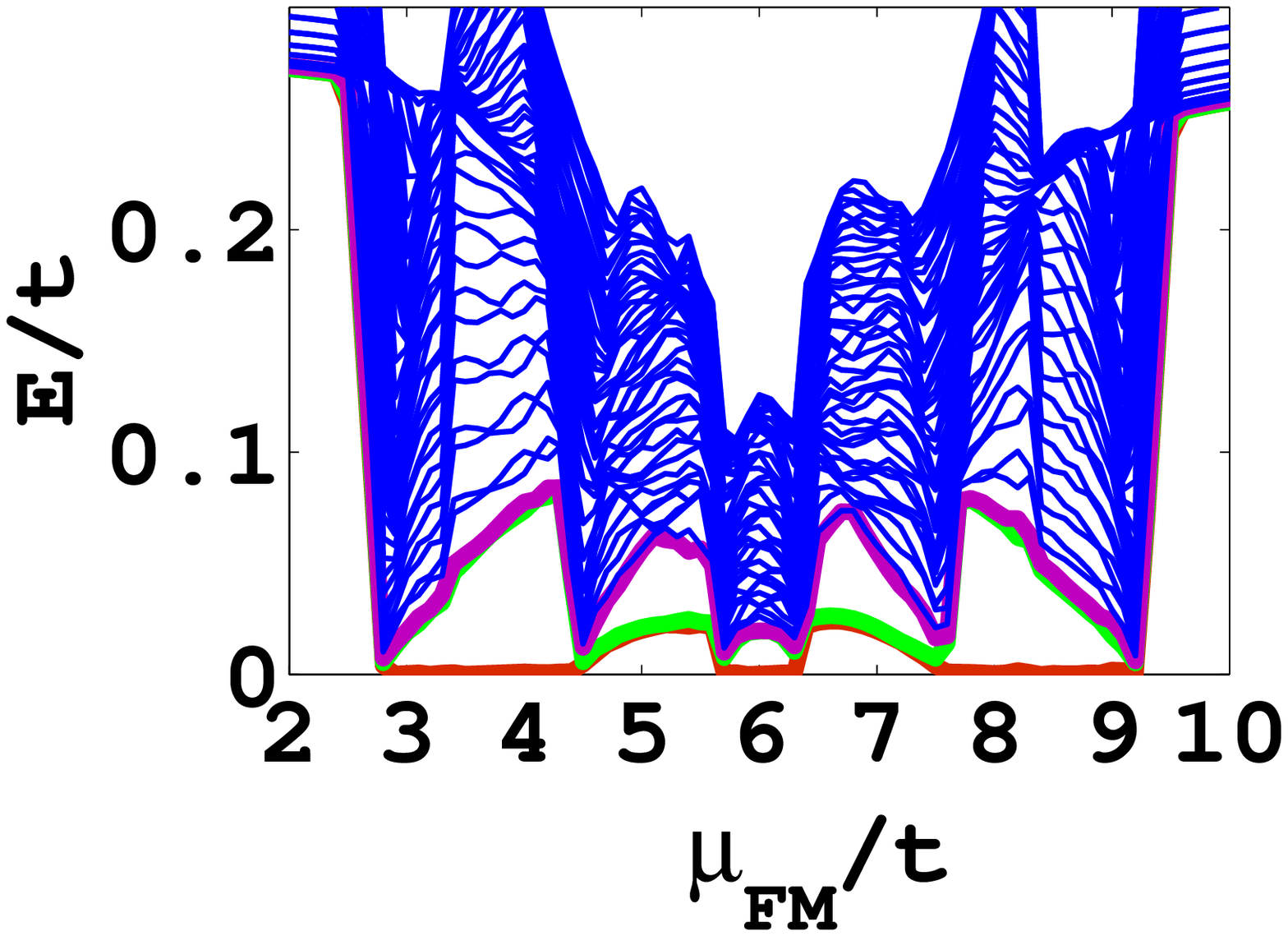}
\includegraphics[scale=0.28]{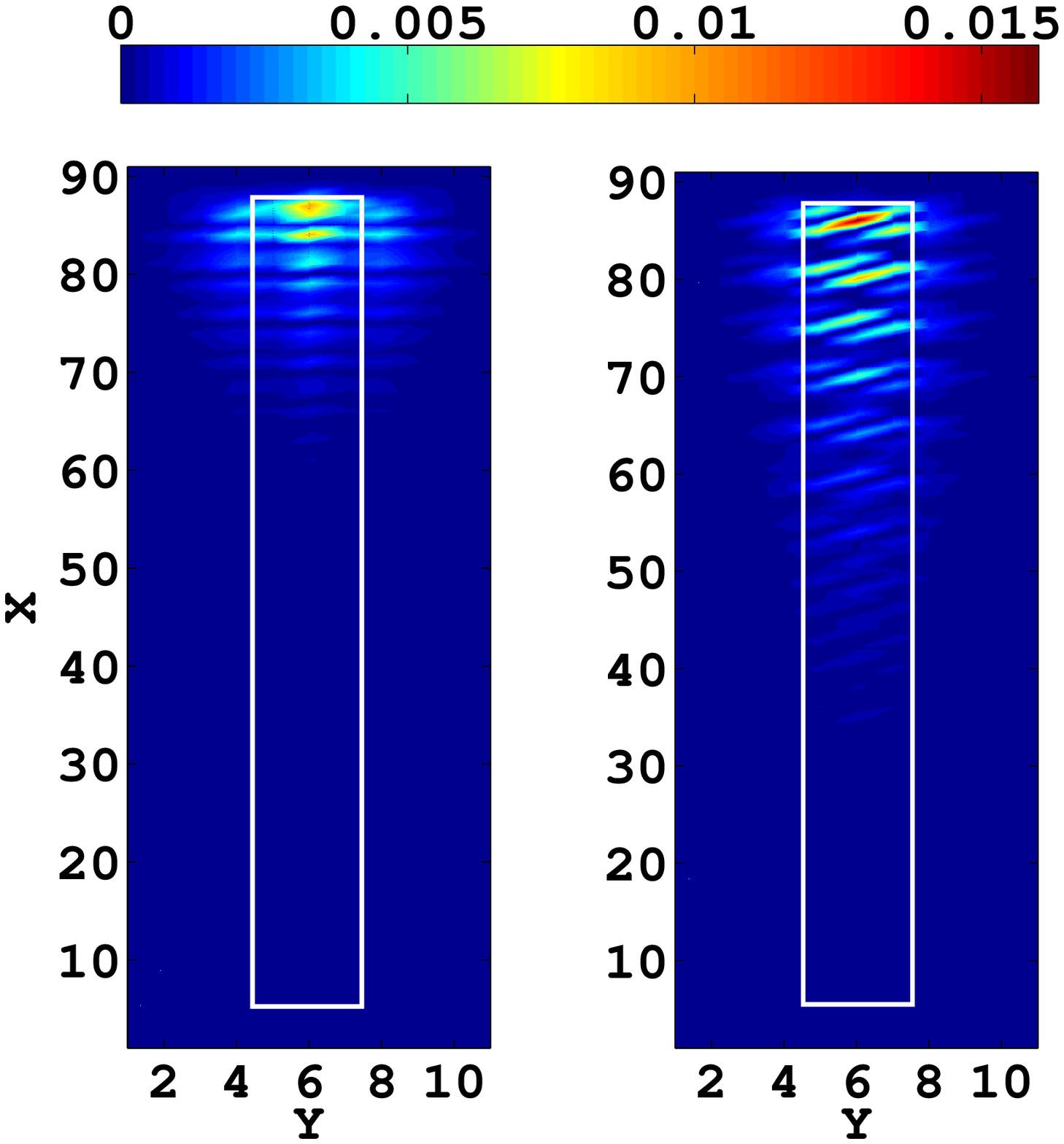}
\caption{\textbf{$\mid$ Quasi-one dimensional wire. a,}
 The finite width $W$ quasi-1D FM wire with eventual tilting of the magnetization.
 Here as well flipping the sign of the perpendicular field and/or of the
 supercurrent and/or that of all in plane fields leaves the results invariant.
\textbf{b,} Typical low-energy quasiparticle spectrum for
$\Delta=1$, $\mu_{SC}=0$, $h_z=6$, $|h_y|=0.8$, $|J|=0.2$,  (all in t units) and
magnetization $\hat{h}_1=\hat{h}_z+\hat{h}_y$, $\hat{h}_2=\hat{h}_z$ and $\hat{h}_3=\hat{h}_z-\hat{h}_y$ for the first, second and third row respectively of this $W=3$ wire.
We observe that a single near zero eigenenergy (red line) emerges when odd numbers of transverse sub-bands
in the wire are partially occupied e.g. near $\mu_{FM}=4$ (1 sub-band) and $\mu_{FM}=6$ (three sub-bands)
as anticipated \cite{Lee}.
For $\mu_{FM}=5$
two transverse sub-bands cross the Fermi level
and the two pairs
of MFs interfere acquiring finite energy.
\textbf{c,} One Majorana mode for $\mu_{FM}=4$ (left) and one for $\mu_{FM}=6$ (right).
The $\mu_{FM}=6$ Majorana mode is less localized because it is protected by a smaller energy gap.
The white rectangle defines the FM wire.} \label{fig:Maj8}
\end{figure}

To illustrate the robustness of these Majorana modes, we extend our discussion to a FM wire of finite width $W$, still small compared to the length $ L$, incorporating a possible tilting of the magnetic moment in the wire as indicated in Figure 3a.
The results of our numerical analysis are shown in Figure 3b and 3c where the finite $W$ corresponds to 3 lattice sites  introducing three bands in the FM wire which are spin split. In Figure 3c is shown only one of the two  MF modes for two topological phases, with $ \mu_{FM}=4 $ (left panel) and $ \mu_{FM} = 6$ (right panel).

The multiple TPTs in Figure 3b yield topologically non-trivial phase in the range of $ \mu_{FM} $, where the FM wire has an odd
number of partially filled bands that could host Cooper pairing, which again is connected with Lifshitz transitions. Additionally we notice that the finite width $W$ allows now for transverse spin triplet pairing, i.e. a field of the type $ g^y_{\bm{i},\bm{j}} \tau_1$ which combines with the component $ g^x_{\bm{i},\bm{j}} \tau_2 $ to a Cooper pair with chiral symmetry (''$ p_x \pm i p_y $'')
 (quartet D in Supplement I). This phase belongs, thus, to the symmetry class  D with a $\Bbb{Z}_2$ topological invariant \cite{Schnyder}. As elaborated in Ref\cite{Lee,Potter},  for $W \lesssim \xi$ where $\xi$ the transverse SC coherence length the D symmetry class yields a pair of zero-energy MFs, if an \emph{odd} number of transverse sub-bands are partially occupied, as is the case in our model calculation.

\begin{figure}[t]
\includegraphics[scale=0.30]{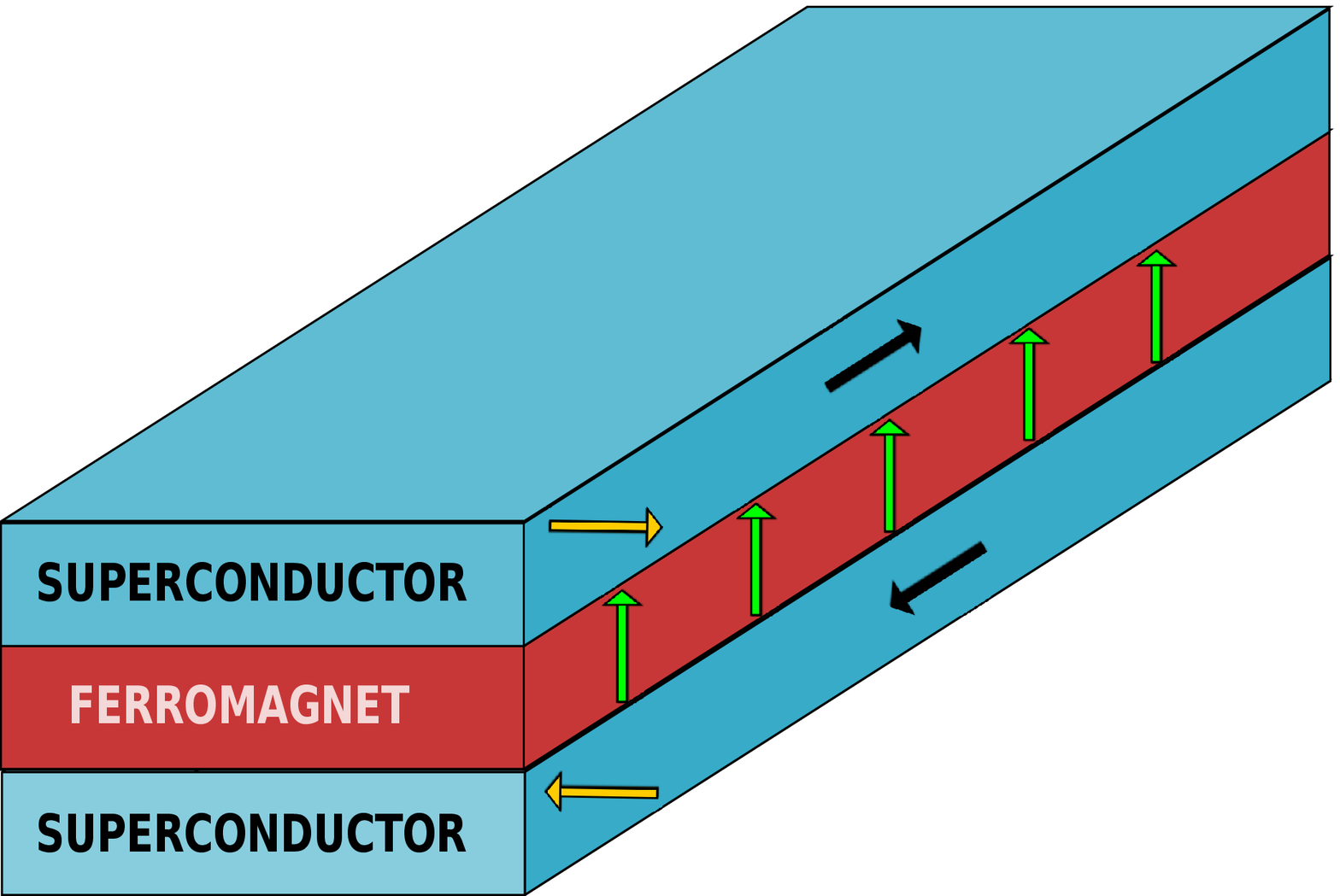}
\includegraphics[scale=0.31]{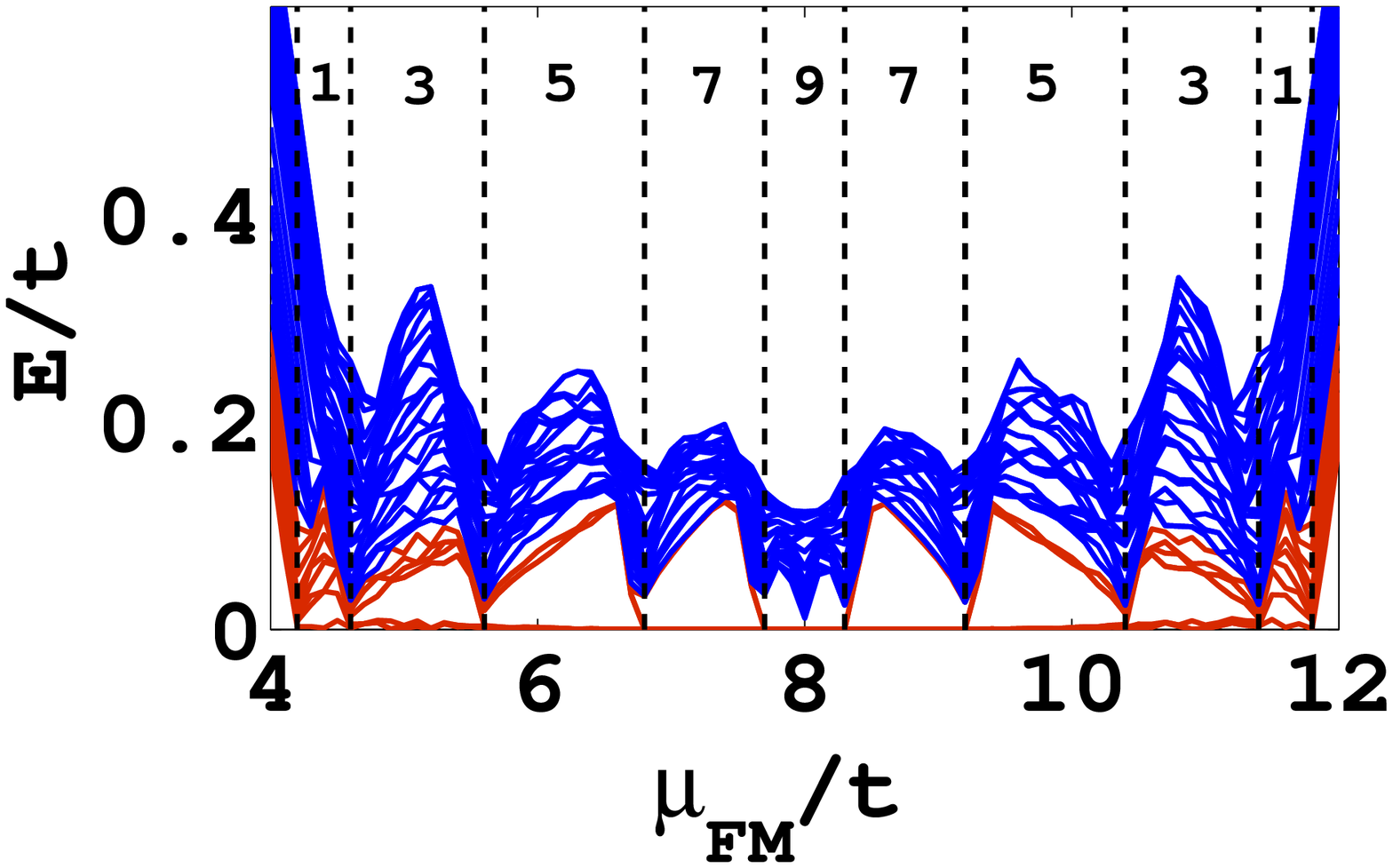}
\includegraphics[scale=0.28]{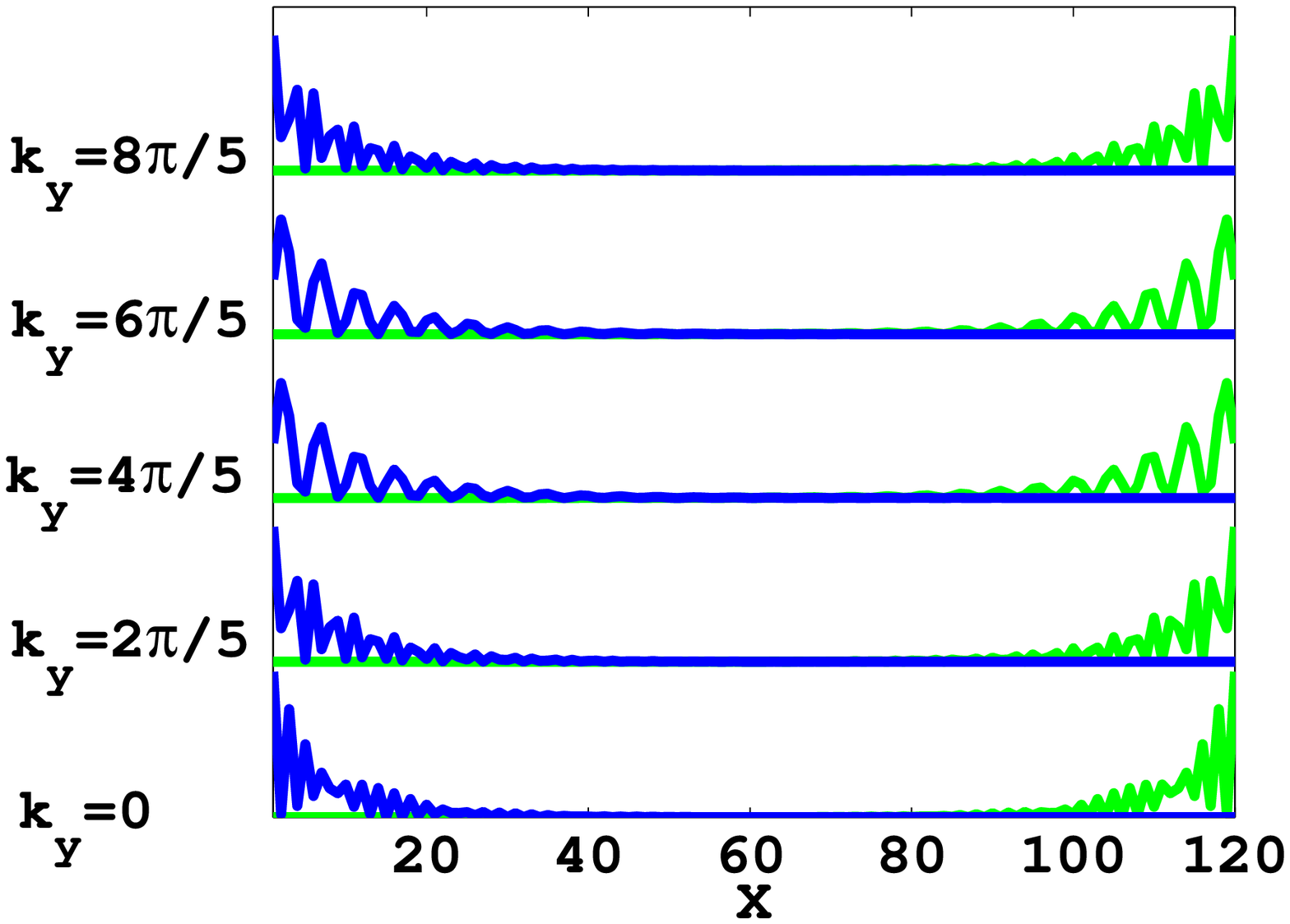}
\includegraphics[scale=0.28]{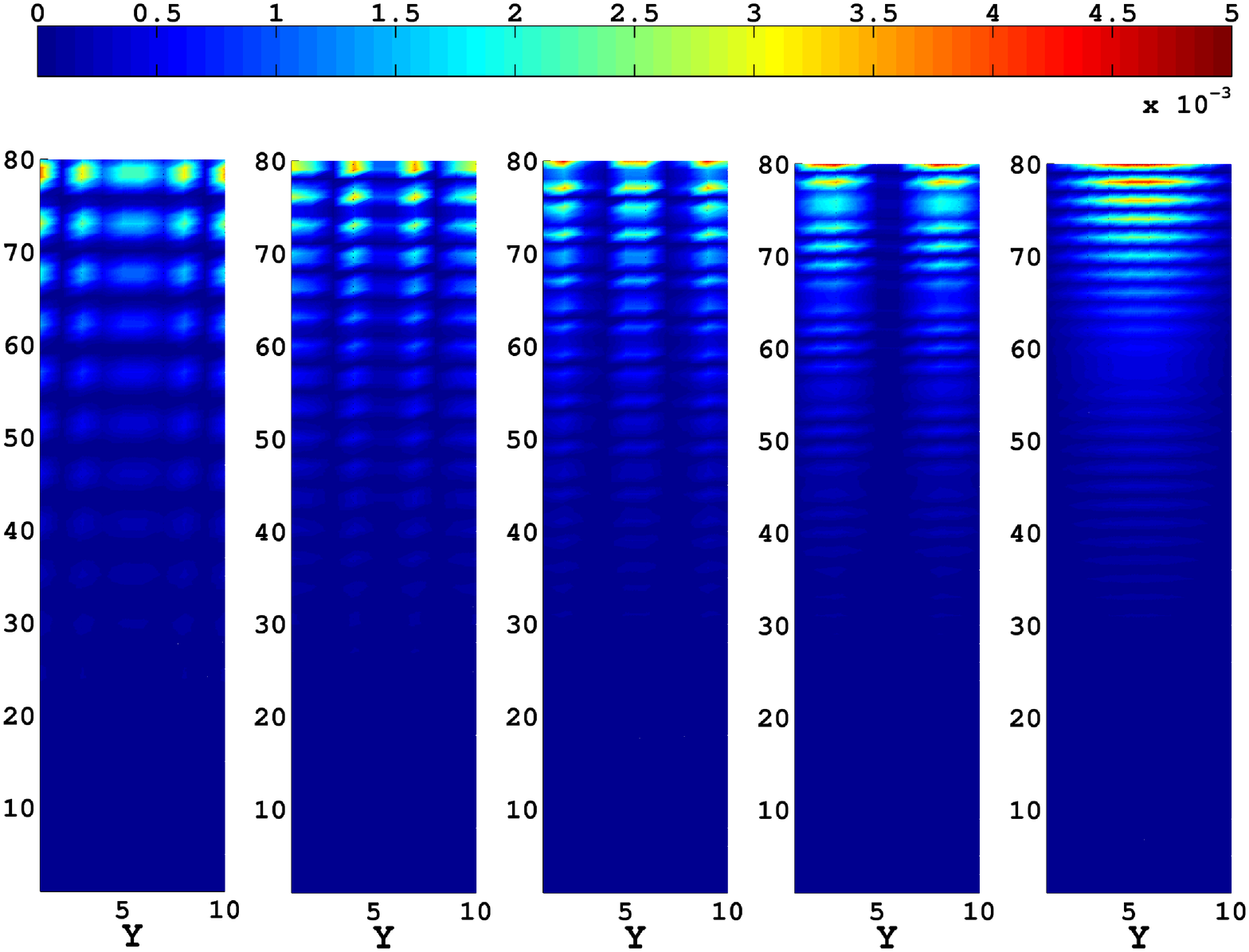}
\caption{\textbf{$\mid$ Layered heterostructure
for controllable generation of Majorana fermions.
 a,} SC/FM/SC trilayer with antiparallel supercurrents (black arrows) and Zeeman  fields
 (yellow arrows), perpendicular to the FM magnetization. Provided supercurrents and in-plane fields
 in the adjacent SC layers remain antiparallel, the signs of fields and currents
 has no influence on the results. Moreover, provided green and yellow Zeeman fields are in
  perpendicular directions, the exact direction of these fields is irrelevant.
 \textbf{b,}
 Typical low-energy quasiparticle spectrum for
$\Delta=4$, $\mu_{SC}=0$,  $h_z=8$, $|h_y|=2$, $|J|=0.6$ and $t_c=0.8$ for the interlayer hopping term all in units normalized
to the in-plane hopping term $t$. Here we have $N_x=120$, $N_y=10$ and periodic boundary conditions along y-axis.
With red lines we denote the branches which pin to zero energy for some
$\mu_{FM}$ values.
Dashed lines indicate the
topological phase transitions
while the numbers on top correspond to the value of the topological invariant $\mathcal{W}$ (see Supplements II and III).
\textbf{c,} The five pairs of Majorana fermions for $\mu_{FM}=6$ corresponding to the
$\mathcal{W}=5$ regime in b,.
\textbf{d,} One Majorana fermion from each of the five Majorana fermion pairs that we obtain
for the same parameters but with open boundary conditions along y-axis.
The system remains manifestly in a BDI symmetry class for both types of boundary conditions.} \label{fig:Maj10}
\end{figure}

After the discussion of MFs in the embedded FM wire we turn to a further related design which might be more
suitable for practical MF engineering. It is important to note that the structure of the device needs not to be one-dimensional,
as in the above device, but that applied currents are sufficient to establish the necessary directionality.
As an example of this kind of device we present here a three-layer structure (see Figure 4a) which consists of a FM layer sandwiched between two conventional SCs. The FM magnetization (green arrows) here points perpendicular to the layer and the
adjacent SC layers carry supercurrents in opposite directions (black arrows) and in-plane Zeeman fields in opposite
directions as well
(yellow arrows).  
The corresponding model Hamiltonian for our numerical analysis is given by

\bea
{\cal H}=\sum_{\bm{i},\bm{j},l,l'}\Psi_{\bm{i},l}^{\dag}\left [ [(\mu_l\tau_3 -\tau_3\bm{h}_l \cdot \tilde{\bm{\sigma}} +\Delta_l\tau_2\sigma_2)\delta_{\bm{i},\bm{j}}  \right . \no \\
\left .+ t_{l}f_{\bm{i},\bm{j}}\tau_3 + \bm{J}_l \cdot \bm{g}_{\bm{i},\bm{j}} ]\delta_{l,l'}+t_{l,l'}\tau_3 \right ]\Psi_{\bm{j},l'}  \label{eq:3DHML}\,,
\eea

\noi where $l$ is a layer index and $t_{l,l'}$ the interlayer hopping term.
The numerical results for such a system of three layers are shown in Figures 4b-d.

Again we see a sequence of TPTs between states involving different number of MF pairs upon changing the chemical potential $ \mu_{FM} $ in the FM layer (Figure 4b).
Although in the particular case demonstrated in Figure 4b only odd number of MFs pairs emerge, in general, also topological phases with even topological invariant can also be reached (Supplement II).

The TPTs of Figure 4b
are understood qualitatively from
an effective Hamiltonian for the FM layer corresponding to our numerical findings
that should exhibit a \emph{parallel rows} structure:
\bea{\cal H}_{FM}^{eff}=\sum_{i,j,\nu,\nu'}\Psi_{i,\nu}^{\dag}[H_{i,j,\nu}^{1D}\delta_{\nu,\nu'}+t'_{\perp}\tau_3\sigma_0\delta_{\nu',\nu \pm 1}\delta_{i,j}]\Psi_{j,\nu'} \label{eq:HML}
\eea
Each row along the x-axis of the FM layer is indexed with $\nu$  and $t'_{\perp}$ is the renormalized transverse inter-row hopping term along the y-axis. The 1D Hamiltonian $H_{i,j,\nu}^{1D}$
has exactly the same form as Eq. \ref{eq::1dRHam}.

The system is translationally symmetric
along the transverse direction when periodic boundary conditions apply
while for open boundary conditions
it only maintains the reflection symmetry.
 In either case the
Hamiltonian Eq. \ref{eq:HML} takes a block diagonal form (Supplement II)
\bea
{ \cal H}_{FM}^{eff}= \sum_{i,j,n}\Psi_{i,n}^{\dag}[H_{i,j,n}^{1D}  + \tau_3\sigma_0\lambda_{n}\delta_{i,j}]\Psi_{j,n}
\eea
where $\lambda_{n}$ are the eigenvalues of  matrix $H_{\perp}=t'_{\perp}\delta_{\nu',\nu \pm 1}$.
Therefore, the system belongs to the $BDI \oplus^{N_y} $  class with the
integer topological invariant $\mathcal{W}=\sum_n \mathcal{W}_n$.
Since $\lambda_{n}$  act as an effective chemical potential which breaks the degeneracy of the 1D sub-systems
\cite{Dumitrescu}, the topological criteria for ${\cal W}_n=1$  are modified accordingly:
$|2t' - \sqrt{h_z^2-\Delta'^2}|< |\mu+t'_{\perp}\lambda_n | <|2t' + \sqrt{h_z^2-\Delta'^2}|$.
For periodic boundary conditions when $t'\approx t'_{\bot}$ and $N_y$ is even, only odd values of ${\cal W}$ are observed as presented in Figure 4b corresponding to Majorana multiplets obeying non-Abelian statistics. For open boundary conditions the residual degeneracy of the transverse bands is lifted and transitions among topological phases with odd and even number of MFs pairs are observed (Supplement II).
We note that the results and discussions presented here are based on a single FM layer,
however this is not a necessary condition as will be discussed in a future work.

To conclude, we have identified \emph{quartets} of fields that are opening
novel extraordinary paths for the
quantum engineering of MFs in conventional SC/FM heterostructures.
No exotic materials with special structures of intrinsic Rashba SOC are needed.
These \emph{quartets} of fields 
have been deliberately discussed here only in the context 
of MF engineering in FM/SC heterostructures. 
Their broader implications 
in a variety of other phenomena will be explored elsewhere. 

\vskip 0.2cm
\noi\textbf{Acknowledgements}

\noi We are grateful to Ali Yazdani for enlightening discussion and insights into his experiment. We also thank
S. Evangelou, P. Kotetes and A.H. MacDonald for stimulating discussions.

\vskip 0.2cm
\noi\textbf{Author Contribution}

\noi G.L. performed the numerical calculations.
G.V. has introduced and proposed the general concept of \emph{quartet rule coupling} applied here
and he supervised the work of G.L. All authors contributed to the development of the project, the analysis
of the results and
the writing of the manuscript.

\vskip 0.2cm
\noi\textbf{Additional Information}

\noi Supplementary information is available in the online version of the paper. Reprints and
permissions information is available online at www.nature.com/reprints. Correspondence
and requests for materials should be addressed to G.V. by mail: varelogi@mail.ntua.gr

\vskip 0.2cm
\noi\textbf{Competing Financial Interests}

\noi The authors declare no competing financial interests.


\begin{thebibliography}{99}

\bibitem{Majorana} Majorana, E. Theoria symmetrica dell' elettrone et del positrone, Nuovo Cimento \textbf{14},
171-184 (1937).

\bibitem{Wilczek} Wilczek, F. Majorana returns, Nat. Phys. \textbf{5}, 614-618 (2009).

\bibitem{Kitaev1} Kitaev, A. Fault-tolerant quantum computation by anyons. Ann. Phys. \textbf{303},
2-30 (2003).

\bibitem{Freedman} Freedman, M. H., Kitaev, A., Larsen, M. J. and Wang, Z. Topological quantum
computation. Bull. Am. Math. Soc. \textbf{40}, 31-38 (2003).

\bibitem{Nayak} Nayak, C., Simon, S. H., Stern, A., Freedman, M. and Das Sarma, S. Non-Abelian
anyons and topological quantum computation. Rev. Mod. Phys. \textbf{80},
1083-1159 (2008).

\bibitem{Alicea} Alicea, Oreg, J.Y. Refael, G. von Oppen, F and Fisher, M.P.A. Non-Abelian statistics and topological quantum information processing in 1D wire networks. Nature Physics, \textbf{7}, 412 (2011).

\bibitem{SigristUeda} Sigrist, M. and Ueda, K. Phenomenological theory of unconventional superconductivity,
Rev. Mod. Phys. \textbf{63}, 239 (1991).

\bibitem{Read} Read, N. and Green, D. Paired states of fermions in two dimensions with breaking
of parity and time-reversal symmetries and the fractional quantum Hall effect.
Phys. Rev. B \textbf{61}, 10267-10297 (2000).

\bibitem{Kitaev} Kitaev, A. Y. Unpaired Majorana fermions in quantum wires. Phys.-Usp. \textbf{44},
131-136 (2001).

\bibitem{Ivanov} Ivanov, D.A. Non-Abelian Statistics of Half-Quantum Vortices in p-Wave Superconductors.
Phys. Rev. Lett. \textbf{86}, 268 (2001).

\bibitem{Fu} Fu, L. and Kane, C. L., Superconducting proximity effect and Majorana fermions
at the surface of a topological insulator. Phys. Rev. Lett. \textbf{100}, 096407 (2008).

\bibitem{LeeProposal} Lee, P. A. Proposal for creating a spin-polarized p$_x$+ip$_y$ state and Majorana
fermions. Preprint at http://arxiv.org/abs/0907.2681 (2009).

\bibitem{Nagaosa} Linder, J., Tanaka, Y., Yokoyama, T., Sudb\"{o}, A. and Nagaosa, N. Unconventional
superconductivity on a topological insulator. Phys. Rev. Lett. \textbf{104},
067001 (2010).

\bibitem{Sau} Sau, J. D., Lutchyn, R. M., Tewari, S. and Das Sarma, S. Generic new platform
for topological quantum computation using semiconductor heterostructures.
Phys. Rev. Lett. \textbf{104}, 040502 (2010).

\bibitem{Oppen} Oreg, Y., Refael, G. and von Oppen, F. Helical liquids and Majorana bound states
in quantum wires. Phys. Rev. Lett. \textbf{105}, 177002 (2010).

\bibitem{AliceaPRB2010} Alicea, J. Majorana fermions in a tunable semiconductor device. Phys. Rev. B
\textbf{81}, 125318 (2010).

\bibitem{Qi} Qi, X-L., Hughes, T. L. and Zhang, S-C. Chiral topological superconductor from
the quantum Hall state. Phys. Rev. B \textbf{82}, 184516 (2010).

\bibitem{Morp} Martin, I. and Morpurgo, A.F.
Majorana fermions in superconducting helical magnets.  Phys. Rev. B \textbf{85}, 144505 (2012).
\bibitem{Flesh}  Kjaergaard, M. W\"{o}lms, K. and Flensberg, K.
Majorana fermions in superconducting nanowires without spin-orbit coupling.
Phys. Rev. B \textbf{85}, 020503(r) (2012).
\bibitem{Kotetes} Heimes, A. Kotetes, P. and G. Sch\"{o}n, G.
Majorana fermions from Shiba states in an antiferromagnetic
chain on top of a superconductor. Phys. Rev. B \textbf{90}, 060507(R) (2014)


\bibitem {Mourik} Mourik, V. \emph{et al.}
 Signatures of Majorana fermions in hybrid superconductor-semiconductor
 nanowire devices. Science \textbf{336}, 1003 (2012).
 \bibitem{ExpVortex} Jin-Peng Xu \emph{et al.}, Experimental detection of a Majorana mode in
 the core of a magnetic vortex inside a toplogical insulator - superconductor Bi$_2$Te$_3$/NbSe$_2$
 heterostructure, Phys. Rev. Lett. {\bf 114}, 017001 (2015).
\bibitem{Nadj} Nadj-Perge, S. \emph{et al.}
Observation of Majorana fermions in ferromagnetic atomic chains on a
superconductor. Science \textbf{346}, 602 (2014).
\bibitem{Bernevig} Li, J. Chen, H. Drozdov, I.K. Yazdani, A. Bernevig, B.A. and MacDonald, A.H.
Topological Superconductivity induced by Ferromagnetic Metal Chains, Phys. Rev. B \textbf{90}, 235433 (2014).
\bibitem{Brydon} Brydon, P.M.R. Das Sarma, S. Hui, H-Y. and Sau, J.D. Topological Yu-Shiba-Rusinov chain from spin-orbit coupling. Phys. Rev. B \textbf{91}, 064505 (2015).
\bibitem{GV}Varelogiannis, G. General rule predicting hidden induced order parameters
and the formation of quartets and patterns of condensates.
Preprint at http://arxiv.org/abs/1305.2976 (2013).
\bibitem{PRLCMR} Varelogiannis, G., Ferromagnetism and Colossal Magnetoresistance from Phase Competition,
Phys. Rev. Lett. \textbf{85}, 4172 (2000).
\bibitem{JoP} Aperis, A. Varelogiannis, G. Littlewood, P.B. and
Simons, B.D., Coexistence of spin density wave, d-wave singlet and staggered ð-triplet superconductivity,
Journal of Physics Cond. Matter \textbf{20}, 434235
(2008).
\bibitem{PRLCeCoIn} Aperis, A. Varelogiannis, G. and Littlewood, P.B.
Magnetic field induced pattern of coexisting condensates
in CeCoIn$_5$, Phys.
Rev. Lett. \textbf{104}, 216403 (2010).
\bibitem{Peng} Peng, Y. Pientka, F. Glazman, L.I. and von Oppen, F. Strong Localization of Majorana End States in Chains of Magnetic Adatoms. Phys. Rev. Lett. \textbf{114}, 106801  (2015).
\bibitem{Schnyder} Schnyder, A.P. Ryu, S. Furusaki, A. and Ludwig, A.W.W. Classification of topological insulators and superconductors in three spatial dimensions. Phys. Rev. B \textbf{78}, 195125 (2008).
\bibitem{Lee} Potter, A.C. and Lee, P.A. Multichannel Generalization of Kitaev's Majorana End States and a Practical Route to Realize Them in Thin Films. Phys. Rev. Lett. \textbf{105}, 227003 (2010).
\bibitem{Potter} Potter A.C. and Lee, P.A. Majorana end states in multiband microstructures with Rashba spin-orbit coupling. Phys. Rev. B \textbf{83}, 094525 (2011).
\bibitem{Dumitrescu}Dumitrescu, E. Stanescu, T.D. and Tewari, S. Hidden-symmetry decoupling of Majorana bound states in topological superconductors. Phys. Rev. B \textbf{91}, 121413(R) (2015).

\end{thebibliography}
\end{document}